\documentclass[conference]{IEEEtran}

\ifCLASSINFOpdf

\else

\fi

\usepackage{amsmath,amssymb}
\usepackage{graphicx}
\usepackage{xspace}
\usepackage{color}
\usepackage{epstopdf}
\usepackage{algorithm}
\usepackage{algorithmic}
\usepackage{setspace}
\usepackage[font=small,labelfont=bf]{caption}
\usepackage{subcaption}
\usepackage{pgfplots, tikzscale}
\usetikzlibrary{plotmarks}
\usetikzlibrary{plotmarks,patterns}
\pgfplotsset{compat=newest}
\usepackage{multirow}
\definecolor{darkpastelgreen}{rgb}{0.01, 0.75, 0.24}
\definecolor{azure}{rgb}{0.0, 0.5, 1.0}
\usepackage[hyphens]{url}
\usepackage{geometry}
\usepackage{fancyhdr}
\graphicspath{ {figures/} }
\newcommand{\systemname}{NL-COMM-Sat\xspace}
\def\thesubsubsectiondis{\unskip\arabic{subsubsection})}

\newcommand\copyrighttext{%
  \footnotesize Accepted for poster presentation at EuCNC \& 6G Summit 2026}

\fancypagestyle{plain}{%
  \fancyhf{}%
  \fancyhead[C]{\copyrighttext}%
}

\begin{document}
\bstctlcite{IEEEexample:BSTcontrol}

\title{\fontsize{18.9}{18.9}\selectfont NL-COMM-Sat: Breaking the Direct Device-to-Satellite Communication Barrier via ``Aggressive'' Non-Orthogonal Transmissions and Non-Linear Processing}

\author{
\IEEEauthorblockN{Konstantinos Nikitopoulos\IEEEauthorrefmark{1}\IEEEauthorrefmark{2} and Chathura Jayawardena\IEEEauthorrefmark{1}}
\IEEEauthorblockA{\IEEEauthorrefmark{1}Wireless Systems Lab, 6GIC, University of Surrey, Guildford GU2 7XH, UK}
\IEEEauthorblockA{\IEEEauthorrefmark{2}Noema Signal Labs, Cambridge, UK}
\vspace{-22pt}
}


\maketitle
\thispagestyle{fancy}

\begin{abstract}
Direct Device-to-Satellite (D2S) communications, which enable direct satellite connectivity with unmodified user equipment (UE), not only expand global coverage but also reshape the evolution of future access networks. However, D2S links face fundamental challenges due to inherently low signal-to-noise ratios (SNRs) and limited spatial multiplexing gains arising from near line-of-sight propagation, both of which severely constrain achievable spectral efficiency.
Despite the lack of spatial multiplexing, this work shows that \emph{aggressive} non-orthogonal transmissions, where multiple users (e.g., four) transmit concurrently over the same frequency resources, even to a single receive antenna, can unlock substantial capacity gains that remain entirely unexploited by existing systems. Realizing these gains in practice, however, requires receiver architectures that, to the best of our knowledge, have not yet been developed.
To this end, we introduce \systemname, an efficient and flexible framework that overcomes this limitation by enabling aggressive non-orthogonal signal transmissions. In contrast to conventional non-orthogonal multiple access (NOMA) schemes, NL-COMM-Sat supports more than two UEs per receive antenna on the same frequency resource. The framework revisits optimal receiver design principles and proposes computationally efficient processing schemes that translate previously unexplored theoretical gains into tangible throughput improvements, even under realistic channel estimation errors and high-mobility Doppler conditions.
Our evaluation shows that NL-COMM-Sat achieves up to a $2\times$ increase in spectral efficiency compared to orthogonal multiple access and NOMA baselines across all considered SNR and Doppler regimes, even with a single-antenna receiver and user speeds of up to 500~km/h.
\end{abstract}
\section{Introduction}

Non-Terrestrial Networks (NTNs) are poised to reshape global connectivity by overcoming the coverage limitations of terrestrial infrastructure, particularly for remote, maritime, and disaster-stricken regions. This vision is driven by the rapid commercial expansion of Low Earth Orbit (LEO) satellite broadband, where direct Device-to-Satellite (D2S)\footnote{We adopt "D2S" (Device-to-Satellite) to emphasize the uplink direction and to avoid ambiguity with "device-to-device," which shares the acronym D2D in the literature.} communication has emerged as a transformative paradigm. Unlike conventional satellite services requiring specialized terminals, D2S leverages standardized NTN frameworks to connect unmodified user equipment (UE) directly to orbiting satellites \cite{SatcomNTN}. Yet as demand for high-throughput connectivity over D2S links grows alongside the number of concurrent UEs \cite{MultibeamMIMO, MIMOsatPHY, D2DsatMIMO}, improving spectral efficiency (SE) under the channel conditions inherent to D2S remains an open and critical challenge.

D2S links are fundamentally constrained by two compounding
impairments: (i) extremely low SNRs and (ii) near line-of-sight (LoS) propagation. The
free-space path loss in LEO links exceeds that of terrestrial
links by roughly $10^{7}\times$, making SNR the dominant
throughput bottleneck. Simultaneously, the LoS nature of the
channel limits spatial multiplexing gains, thereby restricting
the capacity benefits typically achieved by MIMO in terrestrial
systems.


In this work, we propose \systemname, a framework that reframes this challenge entirely. First we show that the low-SNR, LoS regime itself harbors a distinct and previously unexplored capacity growth opportunity, one that only ``aggressive'' non-orthogonal transmissions are uniquely positioned to exploit. 
Specifically, we show that serving more than two UEs per receive antenna, beyond the limits of traditional NOMA, or, equivalently, operating at overloading factors above 200\%  within a single frequency element, unlocks capacity gains that are inaccessible to both orthogonal schemes and conventional two-user NOMA. Realizing these gains requires receiver architectures capable of handling such aggressive overloading and to the best of our knowledge, no existing design meets this need. We therefore introduce novel processing architectures with distinct performance–complexity trade-offs, grounded in optimal receiver principles, that for the first time make the theoretical gains of aggressive non-orthogonal transmissions practically achievable. Importantly, we show that these architectures remain robust to channel estimation errors and Doppler spreads caused by the high orbital velocities of LEO satellites, impairments that are particularly severe at low SNR and that would degrade conventional approaches, establishing \systemname as a viable and practical solution for next-generation D2S communications. Notably, \systemname achieves up to a twofold increase in spectral efficiency while accounting for the reference signal overhead, and while remaining robust against both high Doppler spreads and channel estimation errors, even at speeds of $500$~km/h.

 \vspace{-3pt}
\section{The unexplored D2S transmission capacity gains} 

We first show that significant theoretical gains can be achieved through aggressively non-orthogonal transmissions under the low-SNR and LoS conditions in D2S links, gains that remain largely unexploited. As shown in Fig.~\ref{fig:cap}, the capacity increases monotonically with the number of concurrently active UEs, even in the single-antenna receiver case. This trend is particularly pronounced in the low-SNR regime and follows directly from the well-known capacity formula
\begin{equation}
C = \log\left(1 + \frac{P_t}{\sigma^2}\sum_{k=1}^{K}|h_k|^2\right),
\end{equation}
where $P_t$ is the transmit power, $K$ is the number of UEs, $h_k$ denotes the channel coefficient of the $k$-th UE, and $\sigma^2$ is the noise variance.


\begin{figure}[!ht]
  \centering
  \includegraphics[width=0.7\columnwidth ]{figures/capacitySNR4.tikz}
  \caption{The sum capacity  as the number of UEs increases for a \textbf{single-antenna receiver}.}
   \label{fig:cap}
   \vspace{-8pt}
\end{figure}

\section{NL-COMM-Sat for Aggressive Non-Orthogonal Transmissions}

The \systemname receiver designs are grounded in optimal receiver processing principles and build on recent advances in non-linear receiver techniques \cite{nikitopoulos2024towards,nlcomm2024,demo_camad}. In particular, \systemname comprises two receiver variants: the NL-COMM variant \cite{nikitopoulos2024towards,demo_camad,fijrn}, a computationally efficient method for optimal soft-output detection, analogous to the otherwise impractical soft sphere decoders \cite{STS}; and the NL-COMM+ variant, which further exploits channel code structure to enable practical maximum a posteriori (MAP) processing.



\begin{figure}[!ht]
  \centering
  \includegraphics[width=0.7\columnwidth ]{figures/SEUEs3.tikz}
  \caption{Achievable SE as the number of UEs increases for a \textbf{single-antenna receiver}. A SNR of 4 dB is assumed. }
   	\label{fig:ratewithUEs}
   \vspace{-8pt}
\end{figure}


Fig.~\ref{fig:ratewithUEs} compares the spectral efficiency (SE) achieved by \systemname against orthogonal multiple access (OMA), code-domain NOMA \cite{SCMA2}, and soft successive interference cancellation (SIC) detection.\footnote{We consider OFDM with a 30~kHz subcarrier spacing at 2~GHz carrier frequency, with a 100~ns delay spread modeled using the 3GPP TDL-A channel profile.} The results confirm that \systemname follows the predicted capacity trends and significantly outperforms both OMA and code-domain NOMA, which is typically limited to at most two UEs due to codebook constraints. Notably, \textbf{NL-COMM} incurs less than $3\times$ the complexity of SIC, while \textbf{NL-COMM+} remains \textbf{ below {$10\times$}}.

An important question is whether the proposed receiver architectures retain their theoretical gains under practical impairments, such as high Doppler spreads and channel estimation errors in high-mobility scenarios. We demonstrate that \systemname consistently outperforms OMA and NOMA baselines, delivering up to a 2× improvement in SE, even after accounting for reference-signal overhead.

 \begin{figure}[!ht]
  \begin{subfigure}{.49\columnwidth}
\centering
\includegraphics[width=0.98\columnwidth]{figures/tp5kmh3.tikz}	
\vspace{-8pt}
\caption{5 km/h}
\label{fig:chanest2_5kmh}
\vspace{-5pt}
\end{subfigure}
\begin{subfigure}{.49\columnwidth}
\centering
\includegraphics[width=0.98\columnwidth]{figures/tp500kmh3.tikz}	
\vspace{-8pt}
\caption{500 km/h}
\label{fig:chanest2_500kmh}
\vspace{-5pt}
\end{subfigure}
\caption{Achievable SE for a single-antenna receiver supporting four UEs with radial speeds of \textbf{(a)} 5 km/h and \textbf{(b)} 500 km/h. Results assume a TDL-A channel with practical channel estimation using 3GPP reference signals.}
	\label{fig:chanest2}
    \vspace{-12pt}
\end{figure}

Specifically, in Fig. \ref{fig:chanest2}, we assess the SE when four UEs are supported by a single-antenna receiver, under practical channel estimation constraints at radial speeds of 5 km/h (Fig. \ref{fig:chanest2_5kmh}) and 500 km/h (Fig. \ref{fig:chanest2_500kmh}).
As shown, \systemname can significantly outperform conventional NOMA and OMA, even after accounting for the reference signal overhead and while being robust against channel estimation impairments, in the presence of high Doppler spreads at low SNRs. 



\bibliographystyle{IEEEtran}
\bibliography{reference}

@ARTICLE{STS, 
 author={Studer, C. and Burg, A. and Bolcskei, H.}, 
 journal={IEEE J. Sel. Areas Commun.}, 
 title={Soft-output sphere decoding: algorithms and {VLSI} implementation}, 
 year={2008}, 
 volume={26}, 
 number={2}, 
 pages={290-300}, 
 ISSN={0733-8716}, 
 month={February},}

@ARTICLE{SatcomNTN,
  author={Sattarzadeh, Ata and Liu, Yun and Mohamed, Abdelrahim and Song, Ruiliang and Xiao, Pei and Song, Zhiqun and Zhang, Haipeng and Tafazolli, Rahim and Niu, Chuanfeng},
  journal={IEEE Access}, 
  title={{Satellite-Based Non-Terrestrial Networks in 5G: Insights and Challenges}}, 
  year={2022},
  volume={10},
  number={},
  pages={11274-11283},
  doi={10.1109/ACCESS.2021.3137560}}

@ARTICLE{MIMOsatPHY,
  author={Heo, Jehyun and Sung, Seungwoo and Lee, Hyunwoo and Hwang, Incheol and Hong, Daesik},
  journal={IEEE Commun. Surv. Tutor.}, 
  title={{MIMO Satellite Communication Systems: A Survey From the PHY Layer Perspective}}, 
  year={2023},
  volume={25},
  number={3},
  pages={1543-1570},
  doi={10.1109/COMST.2023.3294873}}

@ARTICLE{MultibeamMIMO,
  author={Schwarz, Robert T. and Delamotte, Thomas and Storek, Kai-Uwe and Knopp, Andreas},
  journal={IEEE Trans. Broadcast.}, 
  title={{MIMO Applications for Multibeam Satellites}}, 
  year={2019},
  volume={65},
  number={4},
  pages={664-681},
  doi={10.1109/TBC.2019.2898150}}

@ARTICLE{D2DsatMIMO,

  author={Bakhsh, Zohre Mashayekh and Omid, Yasaman and Chen, Gaojie and Kayhan, Farbod and Ma, Yi and Tafazolli, Rahim},

  journal={IEEE Commun. Surv. Tutor.}, 

  title={{Multi-Satellite MIMO Systems for Direct Satellite-to-Device Communications: A Survey}}, 

  year={2025},

  volume={27},

  number={3},

  pages={1536-1564},

  doi={10.1109/COMST.2024.3449430}}

@ARTICLE{SCMA2,
  author={Rebhi, Manel and Hassan, Kais and Raoof, Kosai and Chargé, Pascal},
  journal={IEEE Open Journal of the Communications Society}, 
  title={{Sparse Code Multiple Access: Potentials and Challenges}}, 
  year={2021},
  volume={2},
  number={},
  pages={1205-1238},
  doi={10.1109/OJCOMS.2021.3081166}}

@article{nikitopoulos2024towards,
  author       = {K. Nikitopoulos and G. N. Katsaros and M. Filo and C. Jayawardena and R. Tafazolli},
  title        = {{Towards Software-Based, MIMO, Open-RAN PHY Architectures with both Linear and Non-Linear Processing}},
  journal      = {IEEE Commun. Mag.},
  year         = {2024},
  volume={62},
  number={8},
  pages={133-139}
}

@misc{nlcomm2024,
  title = {{NL-COMM}},
 howpublished = {\url{https://nl-comm.com}},
 note         = {(accessed on 15 March 2026)}
}

@INPROCEEDINGS{demo_camad,
  author={Jayawardena, Chathura and Filo, Marcin and Katsaros, George N. and Nikitopoulos, Konstantinos},
  booktitle={Proc. of IEEE CAMAD}, 
  title={{NL-COMM: Demonstrating Gains of Non-Linear Processing in Open-RAN Ecosystem}}, 
  year={2024},
  volume={},
  number={},
  pages={1-2},
  doi={10.1109/CAMAD62243.2024.10943001}}

@Article{fijrn,
AUTHOR = {Jayawardena, Chathura and Katsaros, George Ntavazlis and Nikitopoulos, Konstantinos},
TITLE = {{NL-COMM: Enabling High-Performing Next-Generation Networks via Advanced Non-Linear Processing}},
JOURNAL = {Future Internet},
VOLUME = {17},
YEAR = {2025},
NUMBER = {10},
ARTICLE-NUMBER = {447},
URL = {https://www.mdpi.com/1999-5903/17/10/447},
ISSN = {1999-5903},
}

\end{document}